\documentclass[fleqn,usenatbib]{mnras}

\usepackage{newtxtext,newtxmath}

\usepackage[T1]{fontenc}

\DeclareRobustCommand{\VAN}[3]{#2}
\let\VANthebibliography\thebibliography
\def\thebibliography{\DeclareRobustCommand{\VAN}[3]{##3}\VANthebibliography}

\usepackage{ulem}

\usepackage{graphicx}	
\usepackage{amsmath}	

\title[FAST GPPS Survey -- V. PSR J1901+0658 in DNS system]{
The FAST Galactic Plane Pulsar Snapshot Survey - V. PSR J1901+0658 in a double neutron star system}

\author[Su et al.]{W.~Q. Su,$^{1,2}$ 
J.~L. Han,$^{1,2,3}$\thanks{E-mail: hjl@nao.cas.cn } 
Z.~L. Yang,$^{1,2}$  
P.~F. Wang,$^{1,2,3}$ 
J.~P. Yuan,$^{4,3}$
C. Wang,$^{1,2,3}$
D.~J. Zhou,$^{1}$ \and
T. Wang,$^{1}$ 
Y. Yan,$^{1,2}$ 
W.~C. Jing,$^{1,2}$
N.~N. Cai,$^{1}$ 
L. Xie,$^{1}$ 
J. Xu,$^{1,3}$ 
H.~G. Wang,$^{5,6}$   
R.~X. Xu,$^{7}$   and 
X.~P. You$^{8}$
\\
$^{1}$National Astronomical Observatories, Chinese Academy of Sciences, Jia-20 Datun Road, ChaoYang District, Beijing
   100012, China\\ 
$^{2}$School of Astronomy and Space Science, University of Chinese Academy of Sciences, Beijing 100049, China\\ 
$^{3}$Key Laboratory of Radio Astronomy and Technology,  Chinese Academy of Sciences, Beijing 100101, China \\   
$^{4}$Xinjiang Astronomical Observatory, Chinese Academy of Sciences, 150 Science 1-street, Urumqi 830011, China \\
$^{5}$Department of Astronomy, School of Physics and Materials Science, Guangzhou University, Guangzhou 510006, Guangdong Province, China \\ 
$^{6}$National Astronomical Data Center, Great Bay Area, Guangzhou 510006, Guangdong Province, China \\ 
$^{7}$Department of Astronomy, Peking University, Beijing 100871, China \\ %
$^{8}$School of Physical Science and Technology, Southwest University, Chongqing 400715, China 
}

\date{Accepted XXX. Received YYY; in original form ZZZ}

\pubyear{2015}

\begin{document}
\label{firstpage}
\pagerange{\pageref{firstpage}--\pageref{lastpage}}
\maketitle

\begin{abstract}
Double neutron star (DNS) systems offer excellent opportunities to test gravity theories. We report the timing results of PSR J1901+0658, the first pulsar discovered in the FAST Galactic Plane Pulsar Snapshot (GPPS) Survey. Based on timing observations by FAST over 5 yr, we obtain the phase-coherent timing solutions and derive the precise measurements of its position, spin parameters, orbital parameters, and dispersion measure. It has a period of 75.7 ms, a period derivative of 2.169(6)$\times 10^{-19}$~s~s$^{-1}$, and a characteristic age of 5.5 Gyr. This pulsar is in an orbit with a period of 14.45 d and an eccentricity of 0.366. One post-Keplerian parameter, periastron advance, has been well-measured as being 0.00531(9) deg~yr$^{-1}$, from which the total mass of this system is derived to be 2.79(7) M$_{\sun}$. The pulsar has the mass upper limit of 1.68 M$_{\sun}$, so the lower limit for the companion mass is 1.11 M$_{\sun}$. Because PSR J1901+0658 is a partially recycled pulsar in an eccentric binary orbit with such a large companion mass, it should be in a DNS system according to the evolution history of the binary system. 
\end{abstract}




\section{Introduction}

Double neutron star (DNS) systems are unique laboratories to test gravity theories in the strong-field regime and study binary evolution. Most DNS systems are identified from radio pulsars, enabling precise measurements of their properties, orbital characteristics, and strong-field effects using pulsar timing. The first DNS system, PSR B1913+16, was discovered by \cite{1975ApJ...195L..51H}. Its orbital decay provided crucial evidence for gravitational waves, confirming a key prediction of general relativity  \citep{1989ApJ...345..434T}. The orbital decay in some compact DNS systems leads to the final merger of two neutron stars, some have been detectable by ground-based gravitational wave detectors, e.g., GW170817 \citep{2017PhRvL.119p1101A}. 

Until now, there have been about 3500 pulsars in the Australia Telescope National Facility (ATNF) Pulsar Catalogue\footnote{\url{http://www.atnf.csiro.au/research/pulsar/psrcat/}} \citep{2005AJ....129.1993M}. However, only 27 of them are in DNS systems (or candidates), and 23 DNS are in the Galactic field (including five DNS candidates: PSRs J1753$-$2240, J1755$-$2550, J1759+5036, J1906+0746, and J2150+3427) with an orbital period in the range from 0.078 to 45 d and an eccentricity in the range from 0.064 to 0.828. Radio pulsars in most DNS systems are partially recycled, with a period in the range from 16.9 to 185 ms and a period derivative in the range from $2.7 \times 10^{-20}$ to $1.8 \times 10^{-17}~\mathrm{s~s^{-1}}$. In general, as long as two post-Keplerian (PK) parameters are measured, the masses of two neutron stars in a DNS system as well as the orbital inclination can be determined \citep[e.g.][]{2004hpa..book.....L}. The measurements of multiple PK parameters certainly enable a precise test of gravity theories. For example \cite{2021PhRvX..11d1050K} achieved the most precise confirmation of the general relativity in the strong-field regime to date by measuring seven PK parameters of the double pulsar system, PSRs J0737$-$3039A and B.

\begin{table*}
\begin{center}
\caption{FAST observations for PSR J1901+0658.}
\begin{tabular}{cccccccc}
\hline\hline
Projects    & PI         &  Obs. dates & FAST Obs. mode\footnote[3]{} & Obs. time & No. of Obs. & No. of channels & No. of Pol.\\
\hline
GPPS Survey & J.~L. Han  & 58563-59499 & SnapShot & 5 min &  3  & 2048 / 4096 & 2 / 4\\ 
            &            &             & Tracking & 10 min&  2  & 4096 & 4  \\
PT2020\_0071& J.~P. Yuan & 59152-59279 & Tracking / SwiftCalibration & 8 min &  4  & 1024 & 4 \\
PT2022\_0047& W.~Q. Su   & 59876-60133 & SwiftCalibration & 5 or 15 min &  22  & 2048 & 4 \\
PT2023\_0143& Z.~L. Yang & 60237-60321 & SwiftCalibration & 5 min &  2  & 2048 & 4 \\
PT2023\_0195& W.~C. Jing & 60353-60386 & TrackingWithAngle & 15 min &  2  & 4096 & 4 \\
\hline
\end{tabular}
\end{center}
\label{obs}
\end{table*}

DNS systems can be formed at the end of the evolution of two massive main-sequence stars \citep[e.g.][]{1991PhR...203....1B}. The general scenario is the following. One of the stars evolves and undergoes a supernova explosion, forming a neutron star. Then the survived binary system evolves to be a high-mass X-ray binary system \citep[e.g.][]{2023pbse.book.....T} until the second star expands and fills its Roche lobe, and experiences the Roche lobe overflow. In this phase, the system becomes unstable and forms a common envelope (CE) so that the orbital angular momentum is lost efficiently and the orbit is shrunk. Afterward, the CE is ejected \citep{1976IAUS...73...75P, 2013A&ARv..21...59I}, and the binary system may survive as a neutron star with a helium star companion. When the helium star evolves and fills its Roche lobe, another phase of the Roche lobe overflow occurs and the NS is spun-up \citep{2002MNRAS.331.1027D,2015MNRAS.451.2123T}. 
Eventually, the second star also undergoes a supernova explosion \citep[ultra stripped supernova,][]{2013ApJ...778L..23T, 2015MNRAS.451.2123T}, producing a second NS if the helium star is massive enough. In this case, the binary system evolves into a DNS system, featuring an old, mildly recycled pulsar and a younger canonical pulsar. They are in an eccentric orbit. An alternative scenario for the double neutron star formation is that the DNS system evolves from two stars with nearly equal masses, avoiding the need for CE evolution with a neutron star \citep{1995ApJ...440..270B,2006MNRAS.368.1742D}. Measuring the properties of DNS systems, such as the pulsar's spin periods, orbital periods, eccentricities, and mass distribution, provides invaluable insights into their complicated unknown evolutionary history \citep[e.g.][]{2017ApJ...846..170T,2018MNRAS.481.4009V}.

The Five-hundred-meter Aperture Spherical radio Telescope \citep[FAST,][]{Nan2006} is currently the most sensitive single-dish radio telescope in the world. We are carrying out the FAST Galactic Plane Pulsar Snapshot (GPPS) Survey \citep{2021RAA....21..107H, 2023RAA....23j4001Z} in the FAST visible sky region of the Galactic latitude range of $\pm10\degr$ with the 19-beam L-band receiver covering the frequency range of 1.0–1.5 GHz. Up to now, the FAST GPPS Survey has discovered 637 pulsars\footnote{\url{http://zmtt.bao.ac.cn/GPPS/}}, including more than 76 transient pulsars \citep{2023RAA....23j4001Z}, and five fast radio bursts \citep{2023MNRAS.526.2657Z}. We have obtained the timing solutions for 30 newly discovered pulsars \citep{2023MNRAS.526.2645S}. Here, we report the phase-coherent timing solution of {\it the first pulsar} discovered in the FAST GPPS Survey, PSR J1901+0658, which is in a binary system with a total mass of 2.79 M$_\odot$, the companion of this millisecond pulsar must be another neutron star.

\begin{figure}
\includegraphics[width=0.46\textwidth]{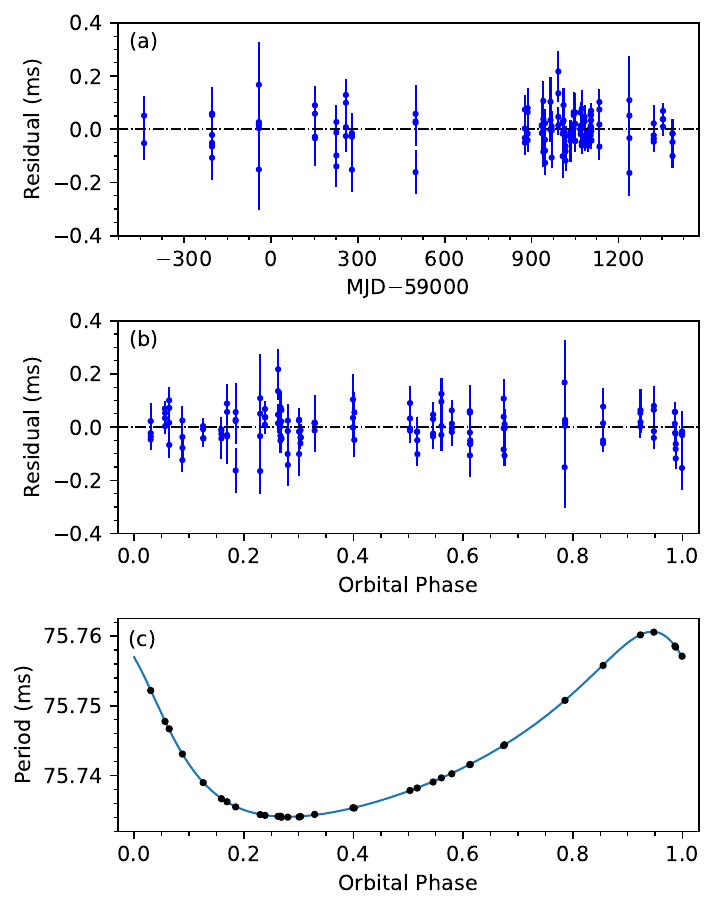}
\caption{
The post-fit timing residuals of PSR J1901+0658 for the phase-connected timing solutions presented in Table~\ref{timingParm_gpps0001}, plotted against MJD in {\it sub-panel (a)} or the orbit phase referenced to the epoch of periastron in {\it sub-panel (b)}. The apparent spin period varies with the orbit phase as shown in {\it sub-panel (c)}. Error bars are included in all plots to represent uncertainties, although some are relatively small and may not be visible. The solid line is the final fitting according to the binary model. 
}
\label{residual}
\end{figure}

\section{FAST observations and results}

PSR J1901+0658 was discovered in the first test observations on 2019 March 21 for the FAST GPPS Survey, and it was given a temporary name PSR J1901+0659g (gpps0001) in \cite{2021RAA....21..107H}. The timing data for this pulsar have been obtained by several FAST observation projects, including PT2020\_0071, PT2022\_0047, PT2023\_0143, and PT2023\_0195, see Table~\ref{obs}. All observations were carried out using the FAST L-band 19-beam receiver in the radio frequency range of 1.0 -- 1.5~GHz. The data are always recorded in search mode, divided into 1024, 2048, or 4096 frequency channels for the band, with a sampling time of 49.152 $\mu$s. For each frequency channel, data were collected for two polarization channels ($XX$ and $YY$) during two GPPS Survey observations, G40.52+0.93\_20190321 and G40.17+1.19\_20200420, but for four polarization channels ($XX$, $YY$, Re[$X^{*}Y$] and Im[$X^{*}Y$]) in the remaining observations.

In the FAST GPPS Survey observations, the ``snapshot 
mode''\footnote{FAST can make observations in different modes with the 19-beam L-band receivers: the ``snapshot mode'' covers a sky area by four pointings of the 19 beams via three beam switches, and these beams are aligned with the Galactic plane; the ``tracking mode'' continuously tracks a position by the central beam (i.e. M01) for some time, with an option for possible data recording for all 19 beams which are aligned with the equator; the ``SwiftCalibration mode'' leads the telescope first to point at a position 10 arcmin away from the target with the on-off calibration signals, and then tracks the target; the ``TrackingWithAngle mode'' first rotates the 19-beam receiver by a given angle from the equator, and then tracks the target by the central beam, with an option for possible data recording for all 19 beams. See the FAST web page \url{https://fast.bao.ac.cn/cms/article/24/} 
for details.} was designed to cover a sky area by using four nearby pointings \citep{2021RAA....21..107H}, with each pointing lasting for 5 min. In the snapshot observation of G40.17+1.19\_20200420, PSR J1901+0658 was detected by the beam M10 in two different pointings, P1M10 and P4M10. We also got it from the beam P3M15 in both G40.52+0.93\_20190321 and G40.52+0.93\_20211012. During the follow-up timing observations, PSR J1901+0658 was tracked for 5 or 15 min each session, utilizing the FAST ``Tracking mode'' or ``SwiftCalibration mode''. Recently, this pulsar has been detected by the beam M11 from two ``TrackingWithAngle mode'' observations, J1901+0708R-11\_20240213 and J1901+0708R-11\_20240317. All these data collected from the FAST GPPS Survey and follow-up observations are used together in this paper to obtain the timing solutions.

With the 19-beam receiver, the FAST has a system temperature of around 22 K. In many observations, we use the periodic calibration noise signals to calibrate the system and polarization. These signals were injected for 2 min (for the 15-min tracking sessions and the snapshot observations) or 40 s (for the 5- and 8-min observation sessions) at the beginning or the end of each observation session.

\begin{figure}
\includegraphics[width=0.46\textwidth]{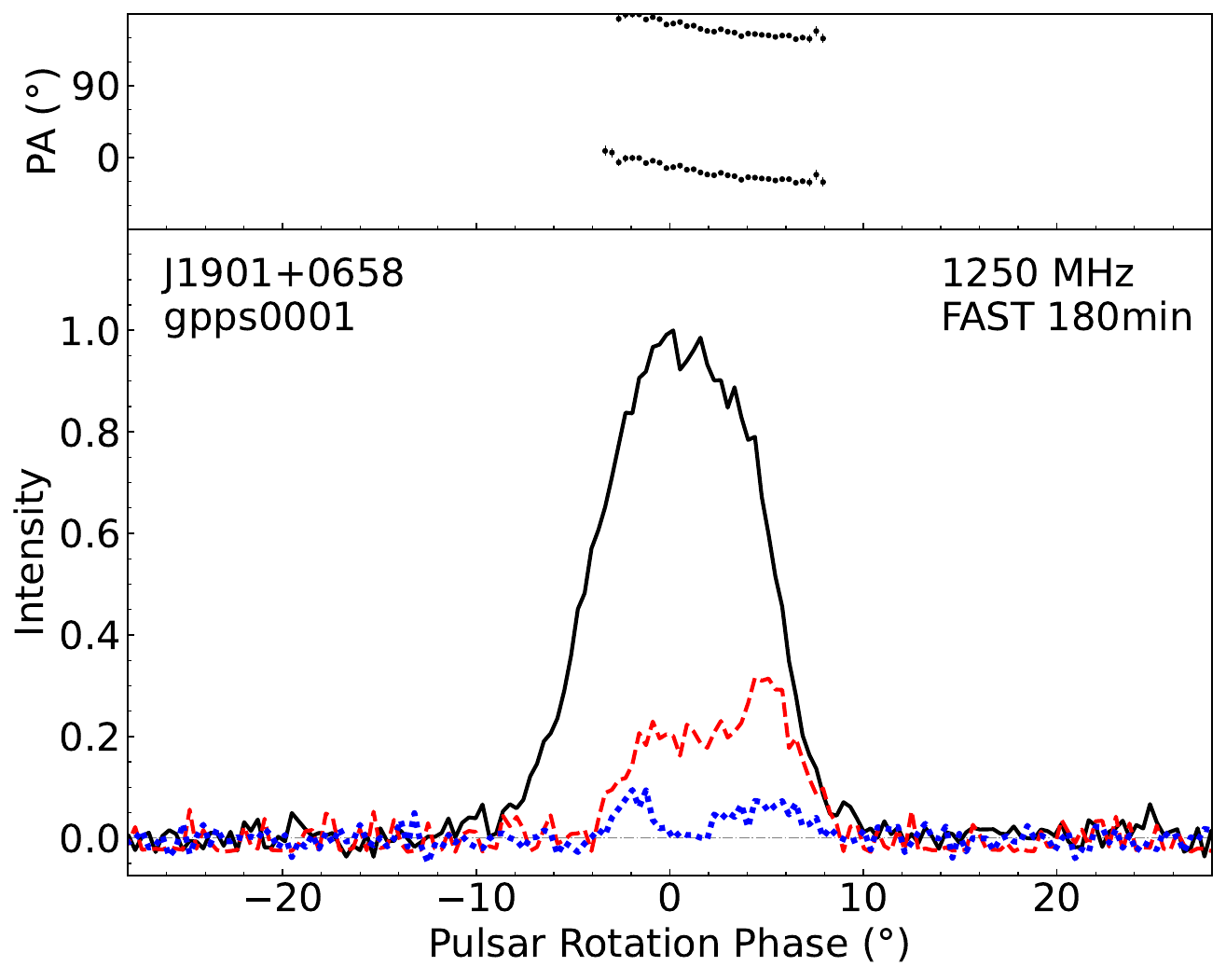}
\caption{Polarization profile of PSR J1901+0658, showing the total intensity (solid line), linear polarization (dashed line), circular polarization (dotted line) in the {\it lower sub-panel}, with the polarization position angle curve displayed in the {\it top sub-panel}. }
\label{profile}
\end{figure}

\begin{table}
\caption{Measured and derived parameters for PSR J1901+0658. The 1$\sigma$ uncertainty of the measured parameters is given in brackets, scaled by the square root of the reduced chi-square.}
\begin{tabular}{ll}
\hline\hline
\multicolumn{2}{c}{General information} \\
\hline
Pulsar name & J1901+0658 \\ 
Data span (MJD) & 58563.1---60386.0 \\  
Number of TOAs & 140 \\
Weighted rms timing residual ($\mu s$) & 52.9 \\
Reduced $\chi^2$ value  & 1.5 \\
Time units  & TCB \\
Solar system ephemeris model  & DE440 \\
Binary model & BT \\
\hline
\multicolumn{2}{c}{Measured parameters} \\ 
\hline
Right ascension, $\alpha$ (hh:mm:ss) &  19:01:22.8512(3) \\ 
Declination, $\delta$ (dd:mm:ss) & +06:58:24.131(6) \\ 
Pulse frequency, $\nu$ (s$^{-1}$) & 13.20239193874(1) \\ 
First derivative of pulse frequency, $\dot{\nu}$ (s$^{-2}$) & $-$3.78(1)$\times 10^{-17}$ \\ 
Dispersion measure, DM (cm$^{-3}$pc) & 125.88(1) \\ 
Epoch (MJD) & 60103.7 \\ 
Orbital period, $P_b$ (d) & 14.4547721(2) \\ 
Epoch of periastron, $T_0$ (MJD) & 59265.530338(8) \\ 
Projected semi-major axis of orbit, $x$ (lt-s) & 32.40204(1) \\ 
Longitude of periastron, $\omega_0$ (deg) & 43.6259(2) \\ 
Orbital eccentricity, $e$ & 0.3662392(5) \\ 
Periastron advance, $\dot{\omega}$ (deg yr$^{-1}$) & 0.00531(9) \\
\hline
\multicolumn{2}{c}{Derived parameters} \\
\hline
Spin period, $P$ (s) & 0.07574385039347(5) \\
Spin period derivative, $\dot{P}$ (s s$^{-1}$)  & 2.169(6)$\times 10^{-19}$ \\
$\log_{10}$(Characteristic age, yr)  & 9.74 \\
$\log_{10}$(Surface magnetic field strength, G)  & 9.61 \\
$\log_{10}$(Edot, ergs/s)  & 31.29 \\
Mass function, $f$ (M$_{\sun}$)  & 0.1748144(1)\\
Total mass, $M_{\rm tot}$ (M$_{\sun}$)  & 2.79(7) \\
DM derived distance, $D_{\rm NE2001}$ (kpc) & 3.8 \\
DM derived distance, $D_{\rm YMW16}$ (kpc) & 3.8 \\
\hline
\multicolumn{2}{c}{Polarization profile parameters} \\
\hline
Pulse width at 50 per cent peak intensities, $W_{50}$ ($\degr$) & 9.8(4) \\
Pulse width at 10 per cent peak intensities, $W_{10}$ ($\degr$) & 15.1(4) \\
Linear polarization percentage, L/I (per cent) & 23(4)\\
Circular polarization percentage, V/I (per cent) & 4(4) \\
Absolute circular polarization percentage, |V|/I (per cent) & 6(3) \\
Rotation measure, RM (rad~m$^{-2}$) &  $-$102(1) \\ 
\hline
\end{tabular}
\label{timingParm_gpps0001}
\end{table}

All FAST observation data of PSR J1901+0658 were folded using the \textsc{dspsr}\footnote{\url{http://dspsr.sourceforge.net/}} \citep{2011PASA...28....1V}, taken the initial position, period, and dispersion measure (DM) obtained from the survey. Subsequently, we used \textsc{pdmp} tool in the \textsc{psrchive} package\footnote{\url{http://psrchive.sourceforge.net}} \citep{2004PASA...21..302H} to refine the optimal period and DM for each observation. The observed periods of different epochs have been Doppler-shifted, indicating the presence of a binary companion. Based on the technique presented in \citet{Bhattacharyya+2008MNRAS.387..273B}, we performed a four-dimensional search to fit the measured spin periods from different epochs, as illustrated in Figure~\ref{residual}(c), employing an eccentric binary orbit model. This allowed us to obtain initial values for binary parameters, including the spin period $P$, orbital period $P_b$, projected semimajor axis of orbit $x$, orbital eccentricity $e$, longitude of periastron $\omega_0$, and epoch of periastron $T_0$. With these first estimated parameters, along with the position obtained during the survey and the optimal DM obtained through \textsc{pdmp}, we proceeded to fold the data into 10-s sub-integrations, each consisting of 1024 phase bins. We also applied \textsc{paz} and \textsc{psrzap} to diminish the radio frequency interference. Subsequently, the folded data was summed into two or four sub-integrations and two frequency channels. Utilizing the data with the highest signal-to-noise ratio (S/N), we generated a noise-free template using \textsc{paas}. Then, times of arrival (TOAs) were obtained using \textsc{pat}.

We then conducted the timing analysis using the \textsc{tempo2} package\footnote{\url{http://bitbucket.org/psrsoft/tempo2}} \citep{2006MNRAS.369..655H}. The initial timing model includes the position, spin frequency $\nu$, derivative of the spin frequency $\dot{\nu}$, $P_b$, $x$, $e$, $\omega_0$, $T_0$, and DM, with the initial $\dot{\nu}$ set to zero. 
Our timing analysis began with these condensed data and the initially fitting $\nu$ and binary parameters. Subsequently, we fitted the new position, $\dot{\nu}$, and DM as necessary. Eventually, we introduced one of the PK parameters, the periastron advance $\dot{\omega}$, into the timing model, which was possible due to the system's significant eccentricity. Using the phase-coherent timing solution, we refolded the data and applied calibration procedures following \cite{2023RAA....23j4002W}. The folded data was again summed into two or four sub-integrations, and all frequency channels were combined. TOAs were recalculated using the new templates derived from integrated pulse profiles, and parameters except DM were refitted to obtain the final ephemeris. The BT binary model \citep{1976ApJ...205..580B}, the JPL DE440 ephemeris, and the TCB units are used in our analysis. We combined all data recorded four polarization channels with a high S/N to obtain the polarization profile for PSR J1901+0658, as shown in Figure~\ref{profile}.

Based on FAST observations for 5 yr, we obtain the timing solution for PSR J1901+0658 as presented in Table~\ref{timingParm_gpps0001}, and the corresponding timing residuals are depicted in Figure~\ref{residual}(a) and (b). The timing solution includes precise measurements of the pulsar's position, spin parameters, orbital parameters, and DM. The weighted residual rms is 52.9 $\mu$s, less than 0.1\% of the pulsar period, indicating a good fit using the timing model. In Figure~\ref{profile}, we present the polarization profiles of PSR J1901+0658 after combining all available data, folded using the timing results. The profile properties are also listed in Table~\ref{timingParm_gpps0001}.

\begin{figure}
\includegraphics[width=0.46\textwidth]{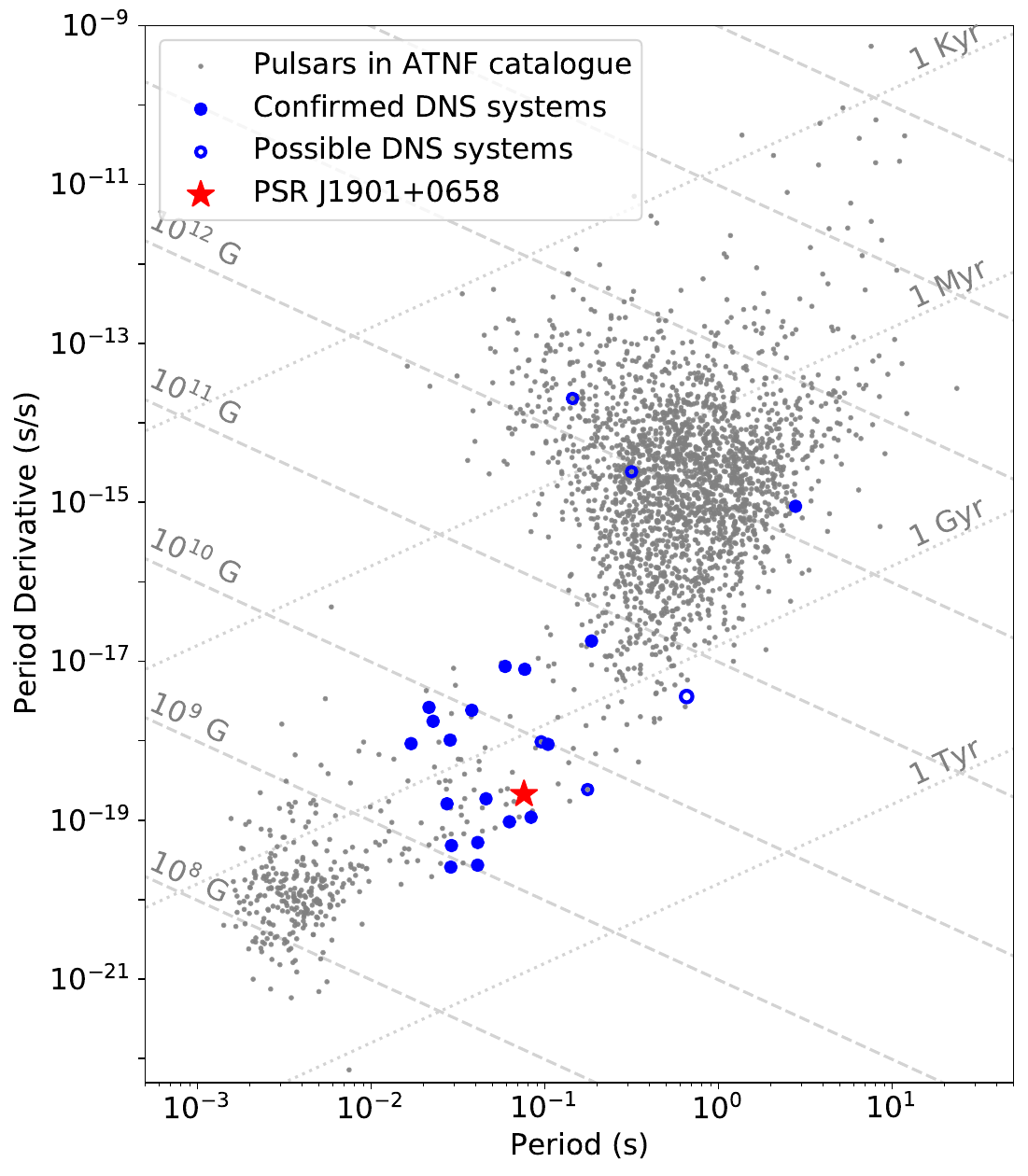}
\caption{The $P-\dot{P}$ diagram for pulsars with indications for PSR J1901+0658 (the star) and other known DNS systems in the Galactic field (dots), DNS candidates (circles). The parameters for all known pulsars (points) are taken from the ATNF Catalogue v1.70 \protect\citep{2005AJ....129.1993M}. Data for three pulsars, PSRs J1018$-$1523, J1208$-$5936, and J2150+3427, are taken from \protect\cite{2023ApJ...944..154S}, \protect\cite{2023A&A...678A.187C}, and \protect\cite{2023ApJ...958L..17W}. The dashed and dotted lines in the background indicate constant surface magnetic field strengths and characteristic ages, respectively.}
\label{PPdot}
\end{figure}

\begin{figure}
\includegraphics[width=0.48\textwidth]{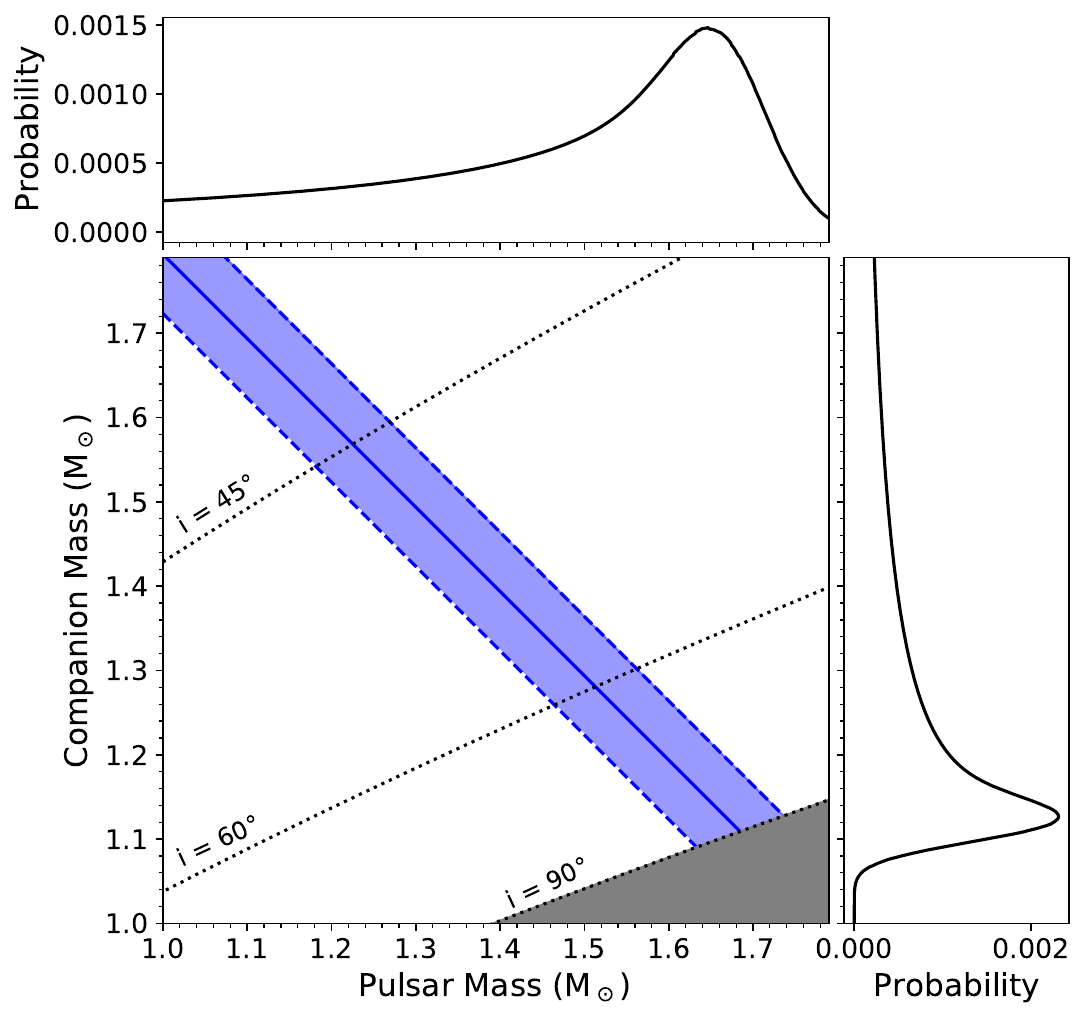}\\
\caption{The mass-mass diagram for PSR J1901+0658, with probability distribution functions for the pulsar mass and companion mass in the {\it top and right sub-panels}. The probability distribution functions are calculated by assuming an isotropic orbital orientation distribution. The shaded area in the bottom right corner is an excluded parameter space defined by the mass function; the area between dashed lines is the possible parameter space constrained by the total mass derived from $\dot{\omega}$, with a uncertainty of $\pm1\sigma$.
}
\label{mass}
\end{figure}

\begin{figure}
\includegraphics[width=0.48\textwidth]{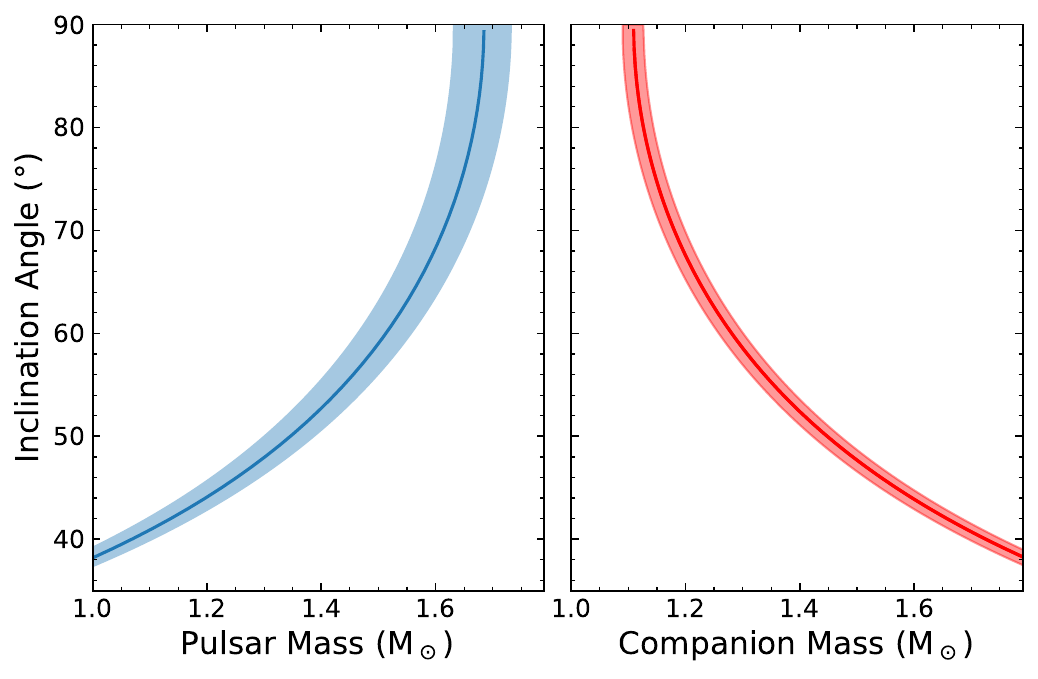}\\
\caption{Possible pulsar mass ({\it left}) and companion mass ({\it right}) versus the possible orbital inclination angle, with the shaded regions for $\pm1\sigma$.}
\label{mass-i}
\end{figure}

\begin{figure*}
\includegraphics[width=0.55\textwidth]{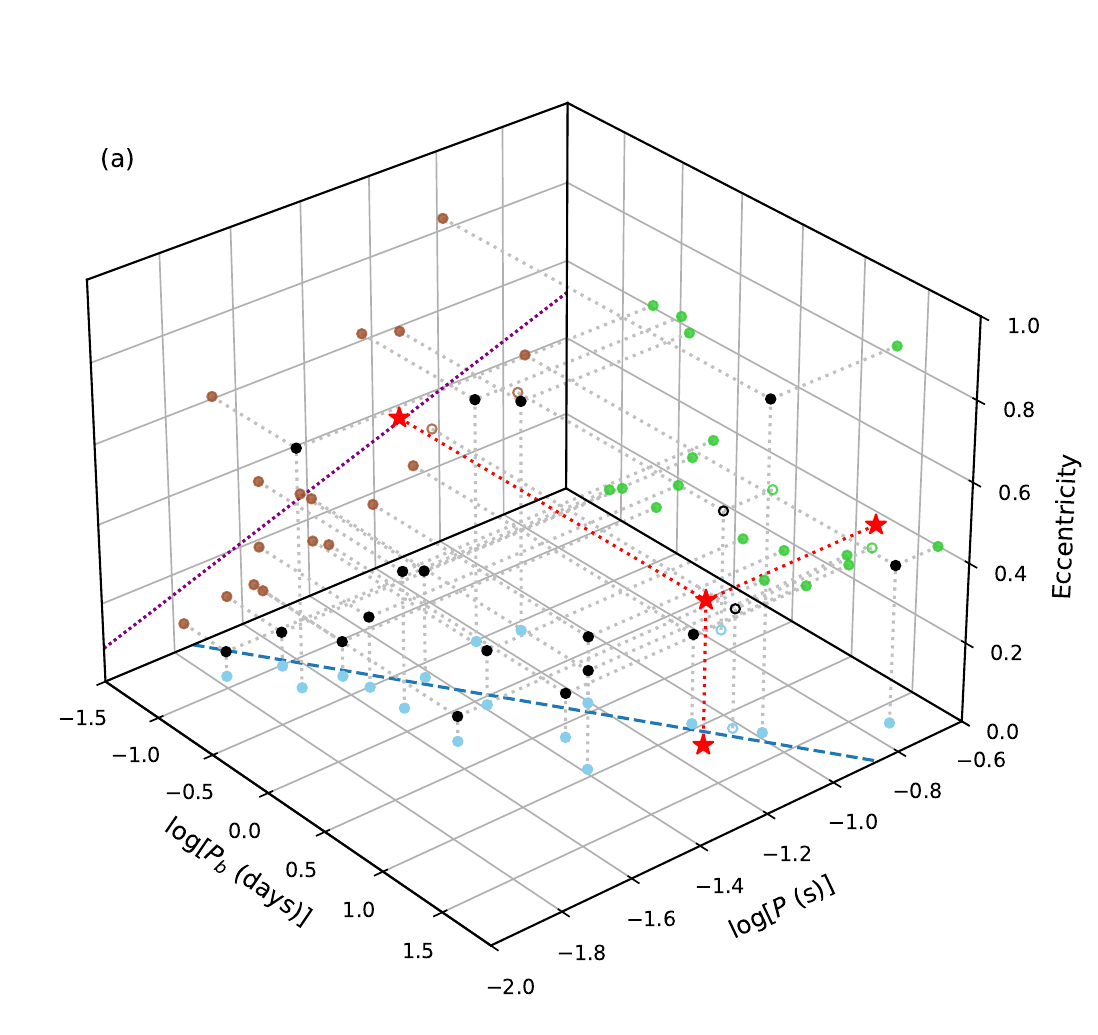}
\includegraphics[width=0.35\textwidth]{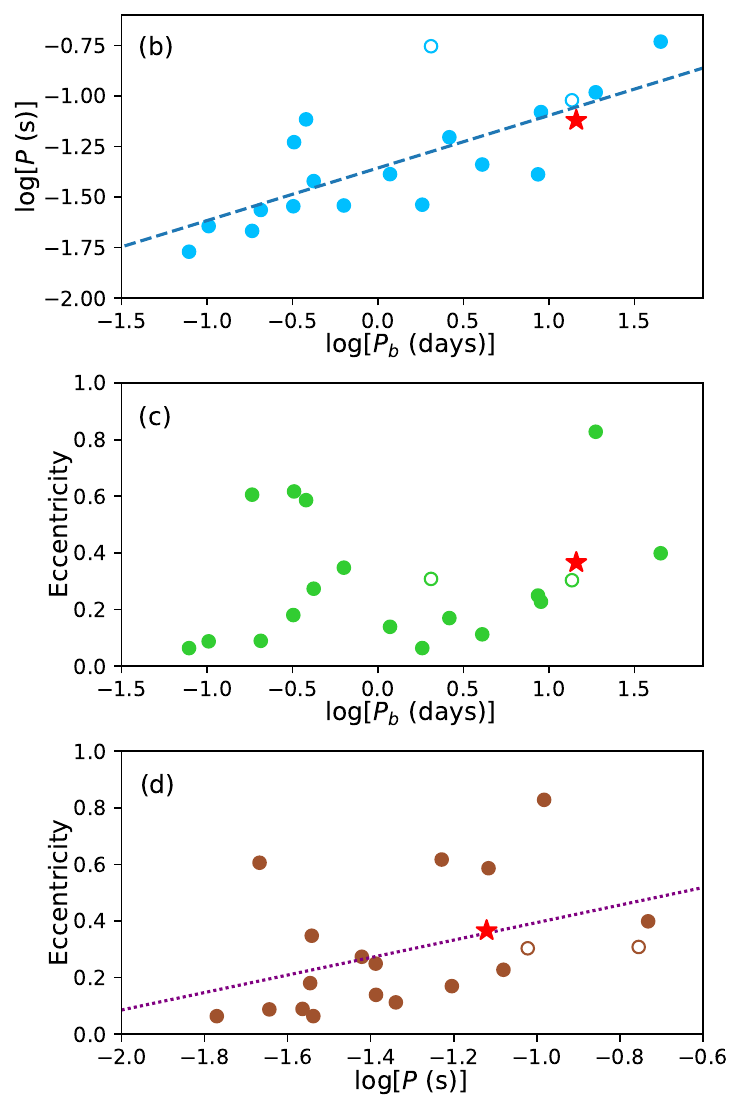}
\caption{The distribution of orbital period, spin period, and eccentricity for recycled pulsars in DNS systems. The orbital period and spin period are well correlated, as seen in {\it sub-panel (b)}, and pulsars in more elliptical orbits tend to be less recycled as seen in {\it sub-panel (d)}. The correlation between orbital period and eccentricity appears very weak as seen in {\it sub-panel (c).} Previously known confirmed DNS systems \citep[data from ][]{2005AJ....129.1993M, 2023ApJ...944..154S, 2023A&A...678A.187C} are denoted by dots, possible DNS systems by open circle \citep{2005AJ....129.1993M}, and PSR J1901+0658 by stars. The light dots represent the projections of known DNS data onto the three planes. The dashed line in the $P_b$-$P$ diagram represents the relation given by \protect\citet{2017ApJ...846..170T}. The dotted line is the best fit to data for spin period and eccentricity. 
}
\label{pb-p-e}
\end{figure*}

\section{Implications and discussions}

PSR J1901+0658 has a period of 75.7~ms and a period derivative of 2.169(6)$\times 10^{-19}$ s s$^{-1}$. It is located in the gap between millisecond pulsars and normal pulsars in the $P$-$\dot{P}$ diagram as shown in Figure~\ref{PPdot}, among typical DNS systems. According to these spin parameters, its characteristic age is estimated to be 5.5 Gyr, the surface magnetic field strength is about 4.1$\times 10^{9}$ G. Its period of several tens of milliseconds, the large age, and the weak magnetic field strength all indicate that it is a partially recycled pulsar, which must have experienced a short accretion process.

According to the Keplerian parameters, PSR J1901+0658 is in a binary system with significant eccentricity. It has an orbital period of 14.45 d, the eccentricity of the orbit is 0.366, and the projected semimajor axis of the orbit is 32.4 lt-s. Using these orbital parameters, we obtain the mass function of
\begin{equation}
    f ( m_{\mathrm{p}} , m_{\mathrm{c}} ) = \frac{(m_{\mathrm{c}} \sin{i})^3}{(m_{\mathrm{p}} + m_{\mathrm{c}})^2} = \frac {4 \pi^{2} x^3} {P_{\mathrm{b}}^2 T_{\sun}} =  0.1748144(1)~\mathrm{M_{\sun}},
\end{equation}
where $T_{\sun} = G M_{\sun} c^{-3} = 4.925490947~\mathrm{\mu s}$, $i$ is the inclination angle of the orbit, and $m_{\mathrm{p}}$ and $m_{\mathrm{c}}$ are the masses of the pulsar and its companion, respectively. 
For pulsars in compact binary systems, the measurement of PK parameters can further constrain the masses of the pulsar and the companion. We have successfully measured one PK parameter for PSR J1901+0658, periastron advance $\dot{\omega}$, which is 0.00531(9) deg yr$^{-1}$. Assuming the periastron advance is caused by the general relativistic effect alone, the rate of periastron advance should be 
\begin{equation}
    \dot{\omega} = 3 T_{\sun}^{2 / 3}\left(\frac{P_{\mathrm{b}}}{2 \pi}\right)^{-5 / 3} \frac{\left(m_{\mathrm{p}}+m_{\mathrm{c}}\right)^{2 / 3}}{1-e^2}.
\end{equation}
The measurement of the periastron advance tells the total mass of this system as being $M_\mathrm{tot} = m_{\mathrm{p}} + m_{\mathrm{c}} =  2.79(7)~M_{\sun}$.

As shown in Figure~\ref{mass}, the masses of PSR J1901+0658 and its companion can be further constrained by the mass function and the total mass deduced from $\dot{\omega}$. Given that $\sin{i} \leq 1$, the shaded region in Figure~\ref{mass} can be ruled out. The total mass is then represented by the blue line in Figure~\ref{mass}, with a shaded area for the $\pm1\sigma$ uncertainty. Considering the upper limit of the inclination angle of $i = 90\degr$, one can get the upper limit for the pulsar mass as being 1.68 M$_{\sun}$, and hence the lower limit for the companion mass as being 1.11 M$_{\sun}$. The pulsar and companion masses depend on the inclination of the orbit, as shown in Figure~\ref{mass-i}, which cannot be constrained yet. The probability distributions for pulsar and companion masses in Figure~\ref{mass} are calculated from the $\dot{\omega}$ by assuming the isotropic orbital orientation distribution. 

PSR J1901+0658 is in a wide orbit with a period of 14.4548~d, other PK parameters are hard to measure. The orbit shrink is only $\dot{P_b}=-10^{-16}\pm 10^{-10}$. For millisecond pulsars in an inclined orbit, the Shapiro delay can be measured even in wide orbits \citep{2024A&A...682A.103G}. For PSR J1901+0658, the Shapiro delay can be measured only for a favourable orbital orientation and high precision TOAs, because it would be only approximately 0.1~ms for a very edge-on orbit ($\sin{i} = 0.999$), while the current TOA uncertainties measured by FAST are in the range of 0.03 to 0.15~ms.
The estimated value of the $\gamma$ value for the Einstein term in the elliptical orbit is 0.01~s, while the available observations can limit the uncertainty to 0.17~s, far from the desired precision. Based on simulations for TOAs with such accuracy, we estimate that more than 30 yr are needed to measure the $\gamma$ at a significance level exceeding 3 $\sigma$.

We attempt but cannot measure the proper motion of PSR J1901+0658  from the current data available. The uncertainties or upper limit of the proper motion in right ascension and declination directions are 6 and 22~mas~yr$^{-1}$, respectively. We cannot detect any variation of the projected semimajor axis, with an upper limit of $4 \times 10^{-13}$~lt-s~s$^{-1}$.

We searched the optical and infrared archival data, the Panoramic Survey Telescope and Rapid Response System (Pan-STARRS) Surveys \citep{2016arXiv161205560C} and Two Micron All Sky Survey \citep[2MASS,][]{2006AJ....131.1163S}, for the corresponding object of the companion, but did not find a white dwarf or a main-sequence companion for this system, which further emphasizes that the companion is another neutron star.

Given that this mildly recycled pulsar is in an eccentric binary orbit with a massive companion, we conclude that PSR J1901+0658 is a member of a DNS system. The second supernova explosion is needed to produce the orbit eccentricity.
As mentioned in Sect.1, for binaries with a wide orbit, during the previous evolution phase of a neutron star with a helium star companion, the helium star undergoes more evolution before filling its Roche lobe and initiating the mass transfer. Consequently, the neutron star has less time for accretion before the second supernova explosion, and therefore it is mildly recycled and has a relatively longer spin period. Therefore, there is a relation between the pulsar spin period $P$ and orbit period $P_b$, as revealed by \cite{2017ApJ...846..170T}:
\begin{equation}
    P = 44  ( P_{\mathrm{b}} / \mathrm{d} ) ^{0.26} \mathrm{~ms}.
\end{equation}
As seen in Figure~\ref{pb-p-e}(b), PSR J1901+0658 is nicely seated in the $P_b$-$P$ diagram for recycled pulsars in DNS systems. 
 
The orbit eccentricity of DNS systems also provides valuable insights into their evolutionary history. The reduced mass transfer tends to be associated with a larger envelope mass and more mass ejection during the second supernova explosion, potentially leading to a higher eccentricity. The neutron star may be kicked during the second supernova explosion that can yield a large eccentricity and a wide orbit \citep{2015MNRAS.451.2123T, 2017ApJ...846..170T}. We check the parameters, $P_b$, $e$, and $P$ in the three dimensional space in Figure~\ref{pb-p-e}. For DNS systems in the Galactic field, as shown in Figure~\ref{pb-p-e}(c), the correlation between orbital period $P_b$ and eccentricity $e$ is very weak. 

Nevertheless, pulsars in more elliptical orbits tend to be less recycled as seen in {\it sub-panel (d)} of Figure~\ref{pb-p-e}, as suggested by \cite{2005ApJ...618L.119F} through the fitting of data from seven DNS systems. It is understandable in the double-star evolution scenario. During the neutron star-helium star binary stage, the less mass transfer to the neutron star results in a larger spin period, while the less mass ejection during the second supernova explosion leads to a smaller eccentricity. The electron-capture supernovae are more symmetric and produce only a smaller kick compared to core-collapse supernovae \citep{2005MNRAS.363L..71D, 2015MNRAS.451.2123T,2024arXiv240105103G}. 

Owing to the wide orbit, PSR J1901+0658 has a merge time of more than $10^{13}$ yr, much longer than the Hubble time. 

\section*{Acknowledgements}
Observation data for this work are taken by using the FAST, a national mega-science facility of China operated by the National Astronomical Observatories, Chinese Academy of Sciences.

The FAST GPPS Survey is a key FAST science project led by JLH. He
coordinated the teamwork and was finally in charge of finishing this paper.
WQS processed all data presented in this paper and drafted the manuscript, under the supervision of JLH. 
PFW developed the processing procedures for pulsar polarization profile and pulsar timing which are extensively used in this paper.
JPY, ZLY and WCJ carried out some of FAST observations presented here. 
Chen Wang feeds all targets for the GPPS observations. 
DJZ, TW, WCJ, YY, LX and  NNC contributed to different aspects of data processing and/or joined many group discussions.
PFW and JX made fundamental contributions to the construction and maintenance of the computation platform.
Other people jointly propose or contribute to the FAST key project.
All authors contributed to the finalization of this paper.

We thank Prof. Youjun Lu for his helpful comments.
The authors are supported by the National Natural Science Foundation of
China (NSFC, Grant Nos 11988101 and 11833009) and the Key Research
Program of the Chinese Academy of Sciences (Grant No. QYZDJ-SSW-SLH021).


\section*{Data Availability}
 
Original FAST observational data will be open resources according to the FAST data 1-yr protection policy. The folded and calibrated profiles in this paper can be found on the webpage: \url{http://zmtt.bao.ac.cn/psr-fast/}.



\bibliographystyle{mnras}
\bibliography{bibfile} 

\begin{thebibliography}{}
\makeatletter
\relax
\def\mn@urlcharsother{\let\do\@makeother \do\$\do\&\do\#\do\^\do\_\do\%\do\~}
\def\mn@doi{\begingroup\mn@urlcharsother \@ifnextchar [ {\mn@doi@}
  {\mn@doi@[]}}
\def\mn@doi@[#1]#2{\def\@tempa{#1}\ifx\@tempa\@empty \href
  {http://dx.doi.org/#2} {doi:#2}\else \href {http://dx.doi.org/#2} {#1}\fi
  \endgroup}
\def\mn@eprint#1#2{\mn@eprint@#1:#2::\@nil}
\def\mn@eprint@arXiv#1{\href {http://arxiv.org/abs/#1} {{\tt arXiv:#1}}}
\def\mn@eprint@dblp#1{\href {http://dblp.uni-trier.de/rec/bibtex/#1.xml}
  {dblp:#1}}
\def\mn@eprint@#1:#2:#3:#4\@nil{\def\@tempa {#1}\def\@tempb {#2}\def\@tempc
  {#3}\ifx \@tempc \@empty \let \@tempc \@tempb \let \@tempb \@tempa \fi \ifx
  \@tempb \@empty \def\@tempb {arXiv}\fi \@ifundefined
  {mn@eprint@\@tempb}{\@tempb:\@tempc}{\expandafter \expandafter \csname
  mn@eprint@\@tempb\endcsname \expandafter{\@tempc}}}

\bibitem[\protect\citeauthoryear{{Abbott} et~al.,}{{Abbott}
  et~al.}{2017}]{2017PhRvL.119p1101A}
{Abbott} B.~P.,  et~al., 2017, \mn@doi [\prl] {10.1103/PhysRevLett.119.161101},
  \href {https://ui.adsabs.harvard.edu/abs/2017PhRvL.119p1101A} {119, 161101}

\bibitem[\protect\citeauthoryear{{Bhattacharya} \& {van den
  Heuvel}}{{Bhattacharya} \& {van den Heuvel}}{1991}]{1991PhR...203....1B}
{Bhattacharya} D.,  {van den Heuvel} E.~P.~J.,  1991, \mn@doi [\physrep]
  {10.1016/0370-1573(91)90064-S}, \href
  {https://ui.adsabs.harvard.edu/abs/1991PhR...203....1B} {203, 1}

\bibitem[\protect\citeauthoryear{{Bhattacharyya} \&
  {Nityananda}}{{Bhattacharyya} \&
  {Nityananda}}{2008}]{Bhattacharyya+2008MNRAS.387..273B}
{Bhattacharyya} B.,  {Nityananda} R.,  2008, \mn@doi [\mnras]
  {10.1111/j.1365-2966.2008.13213.x}, \href
  {https://ui.adsabs.harvard.edu/abs/2008MNRAS.387..273B} {387, 273}

\bibitem[\protect\citeauthoryear{{Blandford} \& {Teukolsky}}{{Blandford} \&
  {Teukolsky}}{1976}]{1976ApJ...205..580B}
{Blandford} R.,  {Teukolsky} S.~A.,  1976, \mn@doi [\apj] {10.1086/154315},
  \href {https://ui.adsabs.harvard.edu/abs/1976ApJ...205..580B} {205, 580}

\bibitem[\protect\citeauthoryear{{Brown}}{{Brown}}{1995}]{1995ApJ...440..270B}
{Brown} G.~E.,  1995, \mn@doi [\apj] {10.1086/175268}, \href
  {https://ui.adsabs.harvard.edu/abs/1995ApJ...440..270B} {440, 270}

\bibitem[\protect\citeauthoryear{{Chambers} et~al.,}{{Chambers}
  et~al.}{2016}]{2016arXiv161205560C}
{Chambers} K.~C.,  et~al., 2016, \mn@doi [arXiv e-prints]
  {10.48550/arXiv.1612.05560}, \href
  {https://ui.adsabs.harvard.edu/abs/2016arXiv161205560C} {p. arXiv:1612.05560}

\bibitem[\protect\citeauthoryear{{Colom i Bernadich} et~al.,}{{Colom i
  Bernadich} et~al.}{2023}]{2023A&A...678A.187C}
{Colom i Bernadich} M.,  et~al., 2023, \mn@doi [\aap]
  {10.1051/0004-6361/202346953}, \href
  {https://ui.adsabs.harvard.edu/abs/2023A&A...678A.187C} {678, A187}

\bibitem[\protect\citeauthoryear{{Dewi}, {Pols}, {Savonije}  \& {van den
  Heuvel}}{{Dewi} et~al.}{2002}]{2002MNRAS.331.1027D}
{Dewi} J.~D.~M.,  {Pols} O.~R.,  {Savonije} G.~J.,   {van den Heuvel} E.~P.~J.,
   2002, \mn@doi [\mnras] {10.1046/j.1365-8711.2002.05257.x}, \href
  {https://ui.adsabs.harvard.edu/abs/2002MNRAS.331.1027D} {331, 1027}

\bibitem[\protect\citeauthoryear{{Dewi}, {Podsiadlowski}  \& {Pols}}{{Dewi}
  et~al.}{2005}]{2005MNRAS.363L..71D}
{Dewi} J.~D.~M.,  {Podsiadlowski} P.,   {Pols} O.~R.,  2005, \mn@doi [\mnras]
  {10.1111/j.1745-3933.2005.00085.x}, \href
  {https://ui.adsabs.harvard.edu/abs/2005MNRAS.363L..71D} {363, L71}

\bibitem[\protect\citeauthoryear{{Dewi}, {Podsiadlowski}  \& {Sena}}{{Dewi}
  et~al.}{2006}]{2006MNRAS.368.1742D}
{Dewi} J.~D.~M.,  {Podsiadlowski} P.,   {Sena} A.,  2006, \mn@doi [\mnras]
  {10.1111/j.1365-2966.2006.10233.x}, \href
  {https://ui.adsabs.harvard.edu/abs/2006MNRAS.368.1742D} {368, 1742}

\bibitem[\protect\citeauthoryear{{Faulkner} et~al.,}{{Faulkner}
  et~al.}{2005}]{2005ApJ...618L.119F}
{Faulkner} A.~J.,  et~al., 2005, \mn@doi [\apjl] {10.1086/427776}, \href
  {https://ui.adsabs.harvard.edu/abs/2005ApJ...618L.119F} {618, L119}

\bibitem[\protect\citeauthoryear{{Gautam}, {Freire}, {Wu}, {Venkatraman
  Krishnan}, {Kramer}, {Barr}, {Bailes}  \& {Cameron}}{{Gautam}
  et~al.}{2024}]{2024A&A...682A.103G}
{Gautam} T.,  {Freire} P.~C.~C.,  {Wu} J.,  {Venkatraman Krishnan} V.,
  {Kramer} M.,  {Barr} E.~D.,  {Bailes} M.,   {Cameron} A.~D.,  2024, \mn@doi
  [\aap] {10.1051/0004-6361/202347836}, \href
  {https://ui.adsabs.harvard.edu/abs/2024A&A...682A.103G} {682, A103}

\bibitem[\protect\citeauthoryear{{Guo}, {Wang}, {Chen}, {Li}, {Ge}, {Jiang}  \&
  {Han}}{{Guo} et~al.}{2024}]{2024arXiv240105103G}
{Guo} Y.-L.,  {Wang} B.,  {Chen} W.-C.,  {Li} X.-D.,  {Ge} H.-W.,  {Jiang} L.,
   {Han} Z.-W.,  2024, \mn@doi [arXiv e-prints] {10.48550/arXiv.2401.05103},
  \href {https://ui.adsabs.harvard.edu/abs/2024arXiv240105103G} {p.
  arXiv:2401.05103}

\bibitem[\protect\citeauthoryear{{Han} et~al.,}{{Han}
  et~al.}{2021}]{2021RAA....21..107H}
{Han} J.~L.,  et~al., 2021, \mn@doi [Research in Astronomy and Astrophysics]
  {10.1088/1674-4527/21/5/107}, \href
  {https://ui.adsabs.harvard.edu/abs/2021RAA....21..107H} {21, 107}

\bibitem[\protect\citeauthoryear{{Hobbs}, {Edwards}  \& {Manchester}}{{Hobbs}
  et~al.}{2006}]{2006MNRAS.369..655H}
{Hobbs} G.~B.,  {Edwards} R.~T.,   {Manchester} R.~N.,  2006, \mn@doi [\mnras]
  {10.1111/j.1365-2966.2006.10302.x}, \href
  {https://ui.adsabs.harvard.edu/abs/2006MNRAS.369..655H} {369, 655}

\bibitem[\protect\citeauthoryear{{Hotan}, {van Straten}  \&
  {Manchester}}{{Hotan} et~al.}{2004}]{2004PASA...21..302H}
{Hotan} A.~W.,  {van Straten} W.,   {Manchester} R.~N.,  2004, \mn@doi [\pasa]
  {10.1071/AS04022}, \href
  {https://ui.adsabs.harvard.edu/abs/2004PASA...21..302H} {21, 302}

\bibitem[\protect\citeauthoryear{{Hulse} \& {Taylor}}{{Hulse} \&
  {Taylor}}{1975}]{1975ApJ...195L..51H}
{Hulse} R.~A.,  {Taylor} J.~H.,  1975, \mn@doi [\apjl] {10.1086/181708}, \href
  {https://ui.adsabs.harvard.edu/abs/1975ApJ...195L..51H} {195, L51}

\bibitem[\protect\citeauthoryear{{Ivanova} et~al.,}{{Ivanova}
  et~al.}{2013}]{2013A&ARv..21...59I}
{Ivanova} N.,  et~al., 2013, \mn@doi [\aapr] {10.1007/s00159-013-0059-2}, \href
  {https://ui.adsabs.harvard.edu/abs/2013A&ARv..21...59I} {21, 59}

\bibitem[\protect\citeauthoryear{{Kramer} et~al.,}{{Kramer}
  et~al.}{2021}]{2021PhRvX..11d1050K}
{Kramer} M.,  et~al., 2021, \mn@doi [Physical Review X]
  {10.1103/PhysRevX.11.041050}, \href
  {https://ui.adsabs.harvard.edu/abs/2021PhRvX..11d1050K} {11, 041050}

\bibitem[\protect\citeauthoryear{{Lorimer} \& {Kramer}}{{Lorimer} \&
  {Kramer}}{2004}]{2004hpa..book.....L}
{Lorimer} D.~R.,  {Kramer} M.,  2004, {Handbook of Pulsar Astronomy}.
~ Vol. 4

\bibitem[\protect\citeauthoryear{{Manchester}, {Hobbs}, {Teoh}  \&
  {Hobbs}}{{Manchester} et~al.}{2005}]{2005AJ....129.1993M}
{Manchester} R.~N.,  {Hobbs} G.~B.,  {Teoh} A.,   {Hobbs} M.,  2005, \mn@doi
  [\aj] {10.1086/428488}, \href
  {https://ui.adsabs.harvard.edu/abs/2005AJ....129.1993M} {129, 1993}

\bibitem[\protect\citeauthoryear{{Nan}}{{Nan}}{2006}]{Nan2006}
{Nan} R.,  2006, \mn@doi [Science in China: Physics, Mechanics and Astronomy]
  {10.1007/s11433-006-0129-9}, \href
  {https://ui.adsabs.harvard.edu/abs/2006ScChG..49..129N} {49, 129}

\bibitem[\protect\citeauthoryear{{Paczynski}}{{Paczynski}}{1976}]{1976IAUS...73...75P}
{Paczynski} B.,  1976, in {Eggleton} P.,  {Mitton} S.,   {Whelan} J.,  eds, ~
  Vol. 73, Structure and Evolution of Close Binary Systems. p.~75

\bibitem[\protect\citeauthoryear{{Skrutskie} et~al.,}{{Skrutskie}
  et~al.}{2006}]{2006AJ....131.1163S}
{Skrutskie} M.~F.,  et~al., 2006, \mn@doi [\aj] {10.1086/498708}, \href
  {https://ui.adsabs.harvard.edu/abs/2006AJ....131.1163S} {131, 1163}

\bibitem[\protect\citeauthoryear{{Su} et~al.,}{{Su}
  et~al.}{2023}]{2023MNRAS.526.2645S}
{Su} W.~Q.,  et~al., 2023, \mn@doi [\mnras] {10.1093/mnras/stad2159}, \href
  {https://ui.adsabs.harvard.edu/abs/2023MNRAS.526.2645S} {526, 2645}

\bibitem[\protect\citeauthoryear{{Swiggum} et~al.,}{{Swiggum}
  et~al.}{2023}]{2023ApJ...944..154S}
{Swiggum} J.~K.,  et~al., 2023, \mn@doi [\apj] {10.3847/1538-4357/acb43f},
  \href {https://ui.adsabs.harvard.edu/abs/2023ApJ...944..154S} {944, 154}

\bibitem[\protect\citeauthoryear{{Tauris} \& {van den Heuvel}}{{Tauris} \& {van
  den Heuvel}}{2023}]{2023pbse.book.....T}
{Tauris} T.~M.,  {van den Heuvel} E. P.~J.,  2023, {Physics of Binary Star
  Evolution. From Stars to X-ray Binaries and Gravitational Wave Sources},
  \mn@doi{10.48550/arXiv.2305.09388.
}

\bibitem[\protect\citeauthoryear{{Tauris}, {Langer}, {Moriya}, {Podsiadlowski},
  {Yoon}  \& {Blinnikov}}{{Tauris} et~al.}{2013}]{2013ApJ...778L..23T}
{Tauris} T.~M.,  {Langer} N.,  {Moriya} T.~J.,  {Podsiadlowski} P.,  {Yoon}
  S.~C.,   {Blinnikov} S.~I.,  2013, \mn@doi [\apjl]
  {10.1088/2041-8205/778/2/L23}, \href
  {https://ui.adsabs.harvard.edu/abs/2013ApJ...778L..23T} {778, L23}

\bibitem[\protect\citeauthoryear{{Tauris}, {Langer}  \&
  {Podsiadlowski}}{{Tauris} et~al.}{2015}]{2015MNRAS.451.2123T}
{Tauris} T.~M.,  {Langer} N.,   {Podsiadlowski} P.,  2015, \mn@doi [\mnras]
  {10.1093/mnras/stv990}, \href
  {https://ui.adsabs.harvard.edu/abs/2015MNRAS.451.2123T} {451, 2123}

\bibitem[\protect\citeauthoryear{{Tauris} et~al.,}{{Tauris}
  et~al.}{2017}]{2017ApJ...846..170T}
{Tauris} T.~M.,  et~al., 2017, \mn@doi [\apj] {10.3847/1538-4357/aa7e89}, \href
  {https://ui.adsabs.harvard.edu/abs/2017ApJ...846..170T} {846, 170}

\bibitem[\protect\citeauthoryear{{Taylor} \& {Weisberg}}{{Taylor} \&
  {Weisberg}}{1989}]{1989ApJ...345..434T}
{Taylor} J.~H.,  {Weisberg} J.~M.,  1989, \mn@doi [\apj] {10.1086/167917},
  \href {https://ui.adsabs.harvard.edu/abs/1989ApJ...345..434T} {345, 434}

\bibitem[\protect\citeauthoryear{{Vigna-G{\'o}mez} et~al.,}{{Vigna-G{\'o}mez}
  et~al.}{2018}]{2018MNRAS.481.4009V}
{Vigna-G{\'o}mez} A.,  et~al., 2018, \mn@doi [\mnras] {10.1093/mnras/sty2463},
  \href {https://ui.adsabs.harvard.edu/abs/2018MNRAS.481.4009V} {481, 4009}

\bibitem[\protect\citeauthoryear{{Wang} et~al.,}{{Wang}
  et~al.}{2023}]{2023RAA....23j4002W}
{Wang} P.~F.,  et~al., 2023, \mn@doi [Research in Astronomy and Astrophysics]
  {10.1088/1674-4527/acea1f}, \href
  {https://ui.adsabs.harvard.edu/abs/2023RAA....23j4002W} {23, 104002}

\bibitem[\protect\citeauthoryear{{Wu} et~al.,}{{Wu}
  et~al.}{2023}]{2023ApJ...958L..17W}
{Wu} Q.~D.,  et~al., 2023, \mn@doi [\apjl] {10.3847/2041-8213/ad0887}, \href
  {https://ui.adsabs.harvard.edu/abs/2023ApJ...958L..17W} {958, L17}

\bibitem[\protect\citeauthoryear{{Zhou} et~al.,}{{Zhou}
  et~al.}{2023a}]{2023RAA....23j4001Z}
{Zhou} D.~J.,  et~al., 2023a, \mn@doi [Research in Astronomy and Astrophysics]
  {10.1088/1674-4527/accc76}, \href
  {https://ui.adsabs.harvard.edu/abs/2023RAA....23j4001Z} {23, 104001}

\bibitem[\protect\citeauthoryear{{Zhou} et~al.,}{{Zhou}
  et~al.}{2023b}]{2023MNRAS.526.2657Z}
{Zhou} D.~J.,  et~al., 2023b, \mn@doi [\mnras] {10.1093/mnras/stad2769}, \href
  {https://ui.adsabs.harvard.edu/abs/2023MNRAS.526.2657Z} {526, 2657}

\bibitem[\protect\citeauthoryear{{van Straten} \& {Bailes}}{{van Straten} \&
  {Bailes}}{2011}]{2011PASA...28....1V}
{van Straten} W.,  {Bailes} M.,  2011, \mn@doi [\pasa] {10.1071/AS10021}, \href
  {https://ui.adsabs.harvard.edu/abs/2011PASA...28....1V} {28, 1}

\makeatother
\end{thebibliography}








\bsp	
\label{lastpage}
\end{document}